\documentclass[aps,amsmath,amssymb,reprint,superscriptaddress]{revtex4-1}
\usepackage{graphicx}

\usepackage{amsmath}
\usepackage{bm}
\usepackage{mhchem}
\usepackage{siunitx}

\usepackage{enumitem} 

\begin{document}
\title{Slow cooling of hot polarons in halide perovskite solar cells}


\date{\today}

\author{Jarvist Moore Frost}
\affiliation{Department of Materials, Imperial College London, Exhibition Road, London SW7 2AZ, UK}
\affiliation{Centre for Sustainable Chemical Technologies and Department of Chemistry, University of Bath, Claverton Down, Bath BA2 7AY, UK}
\email{jarvist.frost@imperial.ac.uk}

\author{Lucy D. Whalley}
\affiliation{Department of Materials, Imperial College London, Exhibition Road, London SW7 2AZ, UK}

\author{Aron Walsh}
\affiliation{Department of Materials, Imperial College London, Exhibition Road, London SW7 2AZ, UK}
\affiliation{Department of Materials Science and Engineering, Yonsei University, Seoul 03722, Korea}
\email{a.walsh@imperial.ac.uk}

\begin{abstract}

Halide perovskites show unusual thermalisation kinetics for above bandgap photo-excitation. We explain this as a consequence of excess energy being deposited into discrete large polaron states. The cross-over between low-fluence and high-fluence `phonon bottleneck' cooling is due to  a Mott transition where the polarons overlap ($n \ge$ \SI{e18}{\per\cm\cubed}) and the phonon sub-populations are shared. We calculate the initial rate of cooling (thermalisation) from the scattering time in the Fr\"ohlich polaron model to be 78 meVps$^{-1}$ for \ce{CH3NH3PbI3}. This rapid initial thermalisation involves heat transfer into optical phonon modes coupled by a polar dielectric interaction. Further cooling to equilibrium over hundreds of picoseconds is limited by the ultra-low thermal conductivity of the perovskite lattice. 

\end{abstract}

\maketitle




A key challenge in the device physics of photovoltaic materials is understanding
where the above bandgap photon energy goes and how to control it.
Thermalisation of `hot' (above bandgap) carriers is normally a fast (\si{\fs}) process in pure crystals.
It is a loss process in photovoltaics and is a major factor underpinning 
the Shockley-Queisser limit for power conversion efficiencies\cite{Shockley1961a}.
To avoid this loss pathway, hypothetical device architectures have been devised
by which these hot carriers can be extracted\cite{green2017energy}.
A fundamental material limit is how \emph{far} the carriers move in the
active photovoltaic layer before cooling to thermal equilibrium.  

There is a growing literature on the kinetics of carrier cooling in halide
perovskites\cite{Price2015,Yang2015,Niesner2016,Zhu2016,Yang2017,Bretschneider2017,Guo2017}.
The behaviour has been linked to a `phonon bottleneck' at high influence, and
more generally to the formation and stability of polaronic charge carriers.  
In addition, it has been established that halide perovskites exhibit low
thermal conductivity, which could be affecting the photophysical processes. 
Thermal conduction in methylammonium lead iodide (CH$_3$NH$_3$PbI$_3$ or MAPI)
is almost as low as a solid-state material can be---the material forms a phonon
glass\cite{Pisoni2014,Whalley2016}.

In this Letter, we consider the microscopic thermal processes in halide
perovskite solar cells underpinning the formation, thermalisation, and cooling
of charge carriers photogenerated from above bandgap illumination. 
We describe how the formation of hot electron and hole charge carriers in the
form of Fr\"ohlich polarons resonates with a sub-population of phonon states
(thermalisation of 78 meVps$^{-1}$), 
which then cool slowly (over hundreds of ps) due to short phonon lifetimes.
We further show that CsPbI$_3$ has a larger thermal conductivity than 
CH$_3$NH$_3$PbI$_3$, which accelerates the kinetics of hot carrier cooling.  


\begin{figure}[h]
\centering
  \includegraphics[width=\columnwidth]{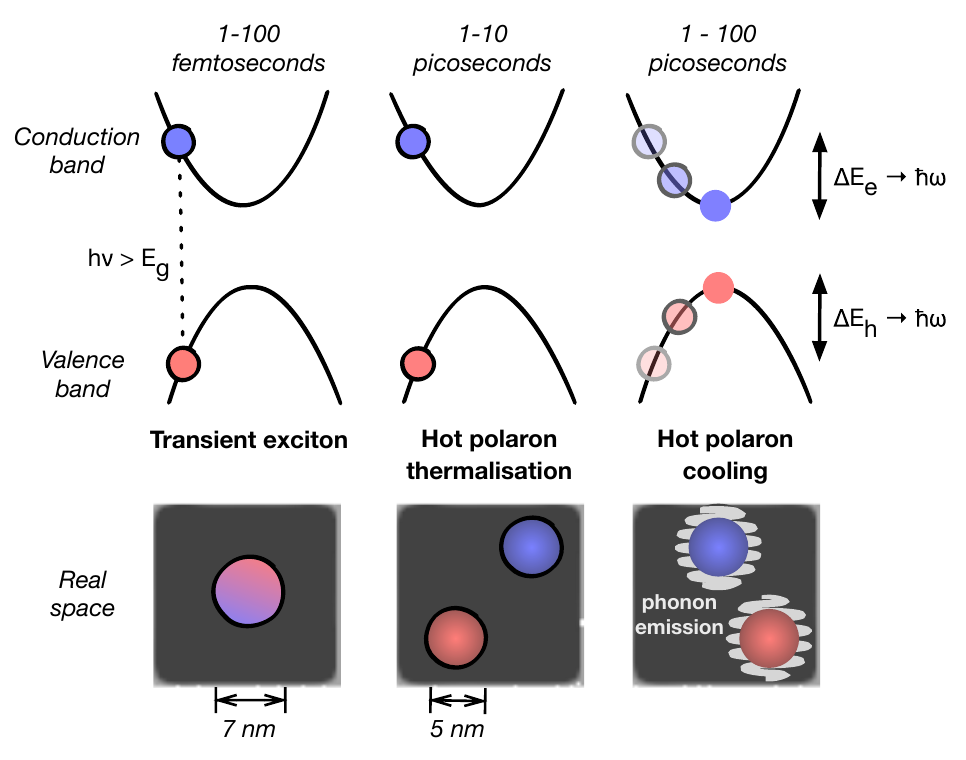}
  \caption{The physical processes involved during the photo-generation of charge carriers which results from above bandgap illumination in a halide perovskite: 
 (i) exciton generation; 
 (ii) exciton dissociation, hot polaron formation and thermalization; 
 (iii) hot phonon relaxation to the band edge limited by the lattice thermal conductivity. 
 Note that there is a distinction between \textit{thermalisation}, which we define as the equilibration with local phonon modes, and \textit{cooling}, which is the equilibration with the extended bulk solid. Together, they form the hot carrier relaxation process.}
  \label{CoolingSchematic}
\end{figure}

\textbf{Measurements and models of carrier cooling.}
Relative to the light intensity generated by a laboratory laser, the sun is dim. 
The charge density maintained by steady-state generation, recombination and extraction of
photogenerated charges under solar irradiation is expected to be around
\SI{e15}{\per\cm\cubed}\cite{Johnston2016}.
Careful control of signal-to-noise is required to reach this low
fluence regime in transient studies. 
The photophysics at higher fluence can be very different
to an operating device under sunlight. 

As well as fluence, there is flexibility in what excitation to pump at. 
In a two-band effective mass model, excitation beyond the bandgap proportionally
results in greater energy electrons and holes.  
However, lead halide perovskites have multiple optically accessible bands.  
Spin-orbit coupling splits the Pb 6p conduction band into levels calculated (by
quasi-particle \textit{GW} theory) at +\SI{1.6}{\eV} and +\SI{3.1}{\eV} above
the valence band maximum\cite{FrostIBSC2016}.  
These values neglect two-particle (excitonic) effects, and electron-phonon renormalization. 
The second transition is observed by spectroscopic ellipsometry as a critical point at 
\SI{2.5}{\eV}\cite{Leguy2016}. 
Exciting well below the experimentally observed second critical point at
\SI{2.5}{\eV} is required to generate a population of hot
carriers in the first conduction band.

The common experimental choice of \SI{400}{\nano\metre} (\SI{3.1}{\eV})
excitation is problematic in terms of interpreting the data. 
We estimate from the partial optical density of states (see
\cite{Leguy2016}, Fig. 4c),  that between 10 and 20\% of this excitation flux
is going into higher conduction bands at \SI{3.1}{\eV}.  
This confuses the analysis, as a combination of (delayed) band-to-band
transitions will overlap with the hot-carrier cooling. 

There is evidence\cite{Price2015,Yang2015,Yang2017} that at high fluence
($n \ge $ \SI{e18}{\per\cm\cubed}), cooling of above-bandgap photo-generated
charges in MAPI is slow ($\tau \approx$ \SI{100}{\pico\second}). 
This has been ascribed to a `phonon bottleneck'\cite{Bockelmann1990} effect. 
Yang et al.\cite{Yang2017} recently studied this high fluence cooling regime in
some detail. 
The existence of a phonon bottleneck even in conventional inorganic quantum
dots is controversial\cite{Li1999}, and requires weak coupling to the fast
dissipating (speed of sound) acoustic vibrational modes. 

A recent transient-absorption microscopy study of polycrystalline MAPI
suggests ballistic transport of the slowly-cooling carriers generated by
excitation at \SI{3.14}{\eV}\cite{Guo2017}. 
%
Similarly, a combined transient-absorption and time-resolved
photo-luminesence study on MAPI\cite{Bretschneider2017} found unusual transient
behaviour when pumping at \SI{3.1}{\eV}. 
They see a direct `cooling', which they associate with a large momentum
transition (i.e. an optical phonon mode) between the Brillouin-zone boundary
and zone centre. 
These unusual data may in part be due to transitions involving
higher conduction bands.

Zhu et al.\cite{Zhu2016} studied the bromine analogue, pumping at modest
fluence ($\approx$ \SI{7e16}{\per\cm\cubed}), $\approx$ \SI{700}{\milli\eV}
above the bandgap. 
No `hot' emission is observed in transient photoluminescence for the inorganic cesium
material, whereas the organic-inorganic materials possess an addition high-energy emission
decaying with a time constant of \SI{160}{\pico\second}.

%

Kawait et al.\cite{Kawai2015} calculated carrier cooling from first-principles via electron-phonon
interactions for \ce{CsPbI3} and bare \ce{PbI3-} octahedra (with a homogeneous
background charge to maintain charge neutrality). 
However, the neglect of spin-orbit coupling to calculate the electronic structure
calls the energy dissipation rate into question, as the conduction
band (Pb 6p) energy, dispersion and degeneracy are significantly altered. 
The electron-phonon coupling was calculated assuming harmonic vibrations, and
thus may further miss the major contribution in highly anharmonic systems such
as the halide perovskites.

All of the transient spectroscopic studies reported so far suggest that a hot
photo-excited state persists in hybrid halide perovskites with a characteristic
cooling time of up to \SI{100}{\pico\second}. 
There are three dynamic processes we need to understand:
1. The photon will first be absorbed into a particular volume of the material (the
exciton, a transient Coulomb bound electron-hole pair).
2. The exciton will then separate into hot carriers (electron and hole polarons),
which will thermalize with the local polaron phonon separation. 
3. The polaron cloud will equilibrate by the transfer of thermal energy to the lattice, 
leading to a cooled charge carrier state. 
These processes are illustrated in Figure \ref{CoolingSchematic}. 
Each of these states can be treated at different levels of theory, from the
microscopic to the mean-field. 
We will discuss them first individually, and then assess the full process.

\textbf{Transient Wannier exciton formation.}
Whether (three dimensional) lead iodide perovskites support an equilibrium population of excitons
(bound electron-hole pairs) is a matter of some experimental debate.
Absorption-based measurements typically indicate the existence of an exciton
state below the bandgap\cite{DInnocenzo2014}, whereas there is no evidence in emission. 
We explain this disagreement as being due to timescale for the measurements.
Absorption probes transient states, whereas emission is sensitive to
a steady state of electrons and holes. 
The difference between the optical dielectric constant ($\epsilon_{\infty} \sim 5$, response 
on a timescale of femtoseconds) and the larger static dielectric
constant ($\epsilon_{0} > 20$, response on a timescale of
picoseconds) means that exciton is transiently stabilised by the optical
dielectric constant. 

Prediction of the exciton state is
a challenge for first-principles electronic structure theory. 
Solution of the Bethe-Salpeter equation (which contains the first order
contribution to electron-hole binding) is computationally demanding, and more
so to achieve convergence. 
The resulting binding energy only considers the response of electronic
excitations (i.e. $\epsilon_{\infty}$). 
The work of Bokdam et al.\cite{Bokdam2016} gives a value for MAPI of
\SI{45}{\milli\eV}.

An effective-mass model of Wannier excitons\cite{Dresselhaus1956} considers the
photo-excited electron and hole to individually be polarons. 
The interaction is statically screened Coulomb interaction of the bare charges. 
This forms a hydrogenic bound state within the nearly-free-electron environment
provided by the band effective masses. 
Simplifying to a single particle system with a reduced effective mass, this is
solved exactly to give a spectrum,   
\begin{equation}
    E_n(k) = 
    -\frac{\mu q^4}{2 \hbar^2 \epsilon^2 n^2} 
    + \frac{\hbar^2 k^2}{2(m_e+m_h)}
\end{equation}
Here, $E_n(k)$ is the energy of state $n$; 
with $k$ the crystallographic momentum. 
For the ground-state of the exciton relative to separated charges ($q$),
$n=1$ and $k=0$ ($\Gamma$ point). 
$\epsilon$ is the dielectric constant
;  $\mu=m_e m_h / (m_e + m_h)$ is the reduced carrier mass.
%
The associated exciton radius (analogous to a Bohr radius) 
is defined as
\begin{equation}
    a_{\mu} = \frac{\epsilon\hbar^2}{\mu q^2}
\end{equation}
Our calculations were cross-checked against CdS\cite{Thomas1959}.

The short timescale (\SI{100}{\fs}) exciton is stabilised by the optical
dielectric constant. 
With values (by QS\textit{GW}\cite{Brivio2014}) of $\epsilon=4.5$, $m_e=0.12$,
$m_h=0.15$, the exciton binding energy is \SI{44.8}{\milli\electronvolt} with
a Bohr radius of $a_{\mu}=$\SI{35.7}{\angstrom}.  
Once the full dielectric response of the lattice occurs
($\epsilon=24.1$), the binding energy reduces to \SI{4}{\milli\eV}, the exciton
orbital expands to an enormous size; the exciton has separated. 
On the timescale of the atomic motion (picoseconds) giving rise to the static
dielectric constant, the exciton decomposes into separate electron and hole
polarons.
The initial hot exciton is transient. 

This model agrees well with a recent study\cite{Luo2017} that measures
the exciton binding energy as \SI{13.5}{\milli\eV}, and associated dephasing
time (which we consider to be the exciton separation) of \SI{1}{\pico\second}. 

\textbf{Large polaron formation.}
A polaron quasi-particle consists of a charge carrier (electron or hole)
wavefunction which has been localised in a dynamically generated potential due
to the polar response of the lattice. 
We recently solved a temperature-dependent Fr\"ohlich polaron model for halide
perovskites\cite{FrostPolarons2017}. 
The calculated Fr\"ohlich polaron coupling constants $\alpha=2.4$ (electron)
and $\alpha=2.7$ (hole) fall in the intermediate coupling regime (defined as
$1<\alpha<6$\cite{Stoneham2001}).
This is because the strong dielectric electron-phonon coupling is balanced
by the light band effective masses, resulting in a strongly interacting large
polaron. 
As such, the continuum (large polaron) theory remains valid, but the effective
mass is strongly renormalised. 

Defining the size of a polaron is difficult. 
The Feynman solution of the Fr\"ohlich Hamiltonian is a simple harmonic
oscillator of the coupled electron-phonon system. 
Schultz\cite{Schultz1959} defines the Feynman polaron radius ($ r_f $) as the standard
deviation of the resulting Gaussian wavefunction ($\psi$), 
\begin{equation} \label{eqn:reducedmass}
    \mu=m(v^2-w^2)/v^2
\end{equation}
\begin{equation} 
    \psi(r)=(\mu v/\pi)^{\frac{3}{4}} exp(-\frac{\mu v r^2}{2})
\end{equation}
\begin{equation} 
    r_f \equiv (<\rho^2>)^{\frac{1}{2}} = (3/2\mu v)^{\frac{1}{2}}
\end{equation}
Here $\rho$ is the density of the wavefunction, $\mu$ is the reduced (effective) mass
of the electron and interacting phonon-cloud, while $v$ and $w$ are internal polaron
parameters characterising the harmonic motion of the polaron. 
The units of $v$ and $w$ are $\hbar\omega$. 
%

\begin{figure}
    \centering
    \includegraphics[width=\columnwidth]{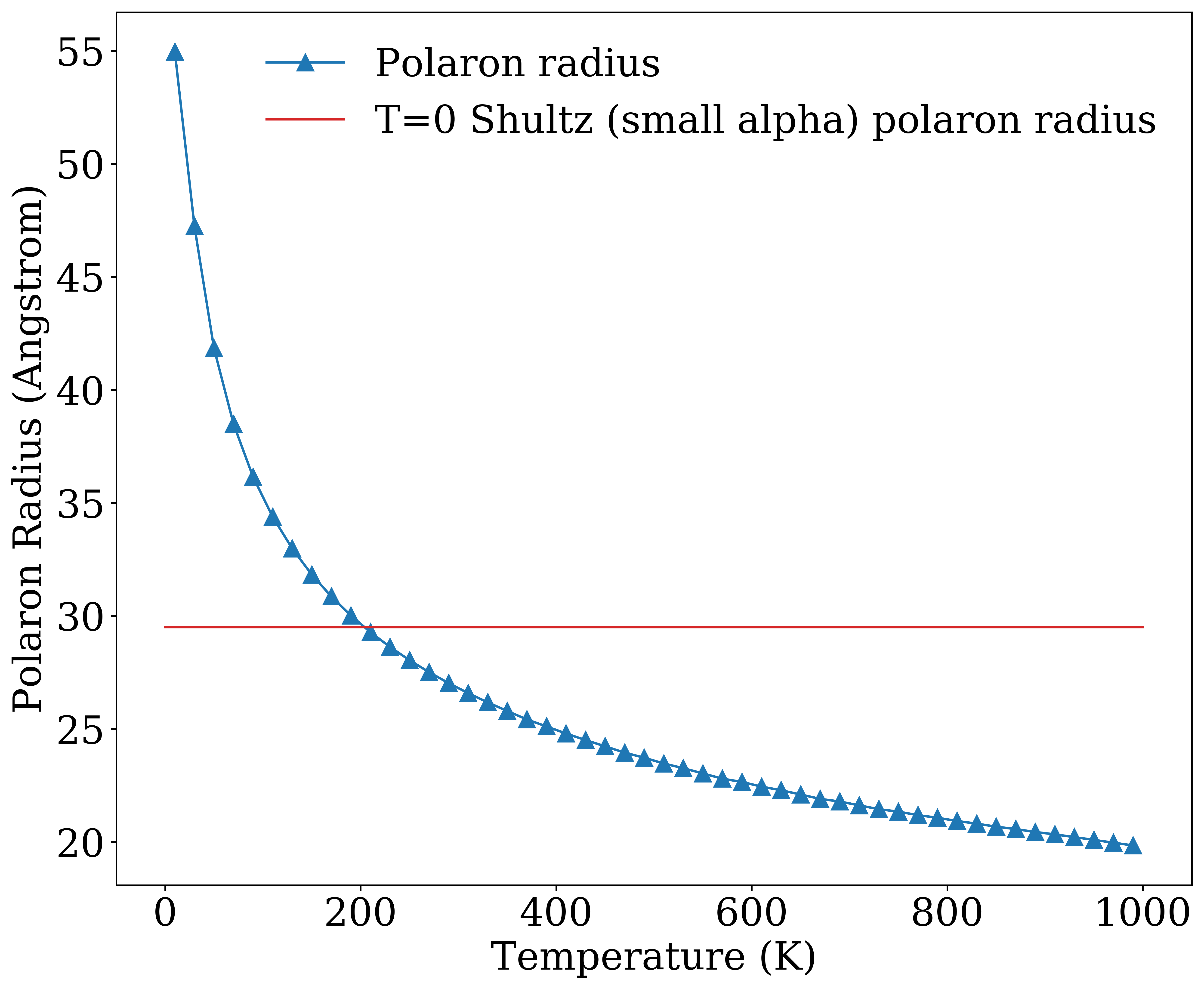}
    \caption{\label{fig:MAPI-electron-radius}
    Temperature-dependent electron ($m_e=0.12$) polaron radius
    (\AA) calculated from a numerical solution to the Fr\"ohlich  polaron model. 
    The horizontal (red) line reprepresents the polaron radius from the common (athermal) small-$\alpha$ approximation.
    }
\end{figure}


Feynman (Eqn. 33 in Ref. \onlinecite{Feynman1955}) provides a zero-temperature variational
problem to solve for $v$ and $w$. 
Assuming $\alpha$ is small, $v=(1+\epsilon)w$, and keeping only linear terms in
$\epsilon$, the integral in the total energy is analytic,
and the variational solution can be approximated as $w=3$,
$v=3+\frac{2}{9}\alpha$\footnote{Following Feynman\cite{Feynman1955}, Eqn. 35
to 36. Thus, $2\mu v = 2m_e \frac{4}{9}\alpha = 2m_e 0.\overline{4} \alpha$. 
(Schultz\cite{Schultz1959} Eqn. 2.5a approximates this as $0.44$.)}. 
We can directly relate the polaron radius to $\alpha$ and $m_e$, 
for MAPI, with $m_e=0.12$, $v=3.53$, the resulting polaron radius is 
$r_f=$\SI{29.5}{\angstrom}.
%
By solving the finite-temperature problem\cite{FrostPolarons2017} with
numeric integration, we can arrive at more accurate values for the
polaron radii of \SI{26.8}{\angstrom} (electron, $v= 19.88$, $w=16.98$) and
\SI{25.3}{\angstrom} (hole, $v=20.11$, $w=16.84$) at \SI{300}{\kelvin}. 
The  internal parameters ($v$,$w$) are roughly linear in temperature. 
The polaron decreases in size as a function of temperature (Figure
\ref{fig:MAPI-electron-radius}). 

We can now estimate at what excitation density the polarons overlap. 
If we define overlap as when the polarons `touch' (i.e. each occupies a cube
with side twice that of the radius), and include a factor of two for capturing
both hole and electron polarons, the density is simply
\begin{equation}
    \rho = V^{-1} = (2(2 r_f)^{3})^{-1} 
\end{equation}
Expressed in standard units, this is \SI{3e18}{\per\cm\cubed}. 
This result provides a simple and direct explanation for the high-fluence
transition to where the carrier cooling is limited by
a bottleneck\cite{Price2015,Yang2015}.  
Namely the polarons are overlapping to the extent that the above-bandgap
thermal energy is shared between overlapping polaron states and cannot dissipate. 
In semiconductor physics, this is a Mott semiconductor-metal transition. 
The phenomenological Mott criterion\cite{Edwards1978} for the polaron overlap 
predicts a density of \SI{4e17}{\per\cm\cubed}. 
These estimates also provide a real-space explanation of the observed lasing threshold of
\SI{e18}{\per\cm\cubed}, in that the electron and hole wavefunctions are
forced into an overlapping (and therefore optically active) configuration.

From the finite-temperature variational model\cite{FrostPolarons2017}, the
polaron free energy is \SI{35.5}{\milli\eV} (electron) and \SI{43.6}{\milli\eV}
(hole) at \SI{300}{\K}.  
The zero-temperature Feynman variational solution gives a phonon occupation
number of $\bar{N}=\frac{\alpha}{2}=1.2$. 
By Bose statistics we would expect this to be $\bar{N}=2.4$ at \SI{300}{\K}.

The excited states (and optical absorption) of polarons have received
considerable theoretical attention.
The path integral solution\cite{Devreese1972} has been recently confirmed for
weak and intermediate coupling by numerical diagrammatic Monte
Carlo\cite{Mishchenko2003}. 
For MAPI, $\alpha=2.4$, we would expect to see a broad absorption band at 1--3 times the
fundamental phonon frequency of $\omega_0$=\SI{2.25}{\THz} (i.e. a broad
photo-excited feature from \SIrange{2.25}{7.75}{\THz}).  
For the intermediate coupling regime, the absorption spectrum is relatively
featureless\cite{Devreese1972,Mishchenko2003}, but a divergence is expected at
frequency $v$, which corresponds to the Franck-Condon (lattice vibration)
transition.  
We estimate $v=20$ (for the electron polaron at \SI{300}{\K}).  
The energy of such a transition ($20\hbar\omega_0$) is \SI{186}{\meV} when
considering the effective phonon mode frequency (\SI{2.25}{\THz}). 
These overtones may be observable in the absorption and emission spectra, but
it is difficult to estimate if they will be discrete features or form an
inhomogeneous broadening. 
They may be misidentified as long lived thermal states. 
The temperature-dependence of the state should follow the
linear dependence of $v$ on temperature. 

The Feynman polaron model includes an inherent dynamic energy exchange.
The coherent oscillation corresponds to energy exchange between electron
and phonon excitations, mediated by the time-retarded dynamic lattice
potential. 
The rate of this oscillation is simply $w$ (Eqn. \ref{eqn:reducedmass}); 
$w=3=$ \SI{6.75}{\THz} with the Feynman
small-$\alpha$ athermal solution, and $w=16.98=$ \SI{38.2}{\THz} with the
\SI{300}{\K} solution.

There are large uncertainties in the above quoted values, due to the
approximations made in these theories, not least the reduction from multiple
phonon branches to a single effective mode following Hellwarth \textit{et
al.}\cite{Hellwarth1999}. 
However, these predictions suggest that large polarons in MAPI may have observable
spectroscopic features that could be used to characterise its internal state. 


\textbf{Hot polaron states.}
We have established that the size of the transient exciton is
commensurate with the polaron state. 
We expect the exciton to quickly (on a timescale of picoseconds) 
decompose into polarons. 
As the bare-band effective-masses in halide perovskites are nearly balanced,
the hole and electron polarons are similar in character and size. 
Without a more detailed physical picture of the process, we assume an
equipartition of the above-bandgap energy ($h\nu > 1.6$ eV for MAPI)  into the
hot hole and electron polaron states. 
Considering excitations up into the near-UV at \SI{4.0}{\eV}, the initial
polaron energy could be as high as \SI{1.2}{\eV}.

An excess carrier energy of \SI{1.2}{\eV} translates by $E=k_B T$ to a single
degree-of-freedom `electron temperature' of \SI{13900}{\K}.
A way of interpreting the high temperatures extracted from transient
experiments is to invert this identity, and calculate amongst how many
microscopic states the excess energy has so far been shared. 
This way an estimate is made of the size of the thermal bath, the subpopulation
of states coupled to the hot carrier. 
Once fully thermalised (local equipartition), this energy will be shared
amongst all accessible phonon states within the polaron. 
In a continuum model, the eventual (full thermalised) polaron temperature
depends on the volume and specific heat capacity ($C_V$) of the polaron. 
%
%
This we can calculate from the phonon density of states. 
Summing over the Bose-Einstein occupied phonon modes for MAPI, we find
a per-unit cell specific heat capacity of \SI{1.25}{\milli\eV\per\K} at
\SI{300}{\K}. 
An electron polaron of radius \SI{26.8}{\angstrom} occupies 360 unit cells of the crystal. 
The maximum initial temperature from considering the above-bandgap energy
(\SI{1.2}{\eV}) being distributed thermodynamically across the inorganic
phonon modes associated with the phonon unit cells is \SI{3}{\kelvin}. 
This temperature seems too small to explain the low-fluence hot-carrier results. 
Instead some mechanism to cause greater confinement, or a reduced effective
specific heat capacity, must be invoked. 

\begin{figure}[h]
\centering
  \includegraphics[width=0.95\columnwidth]{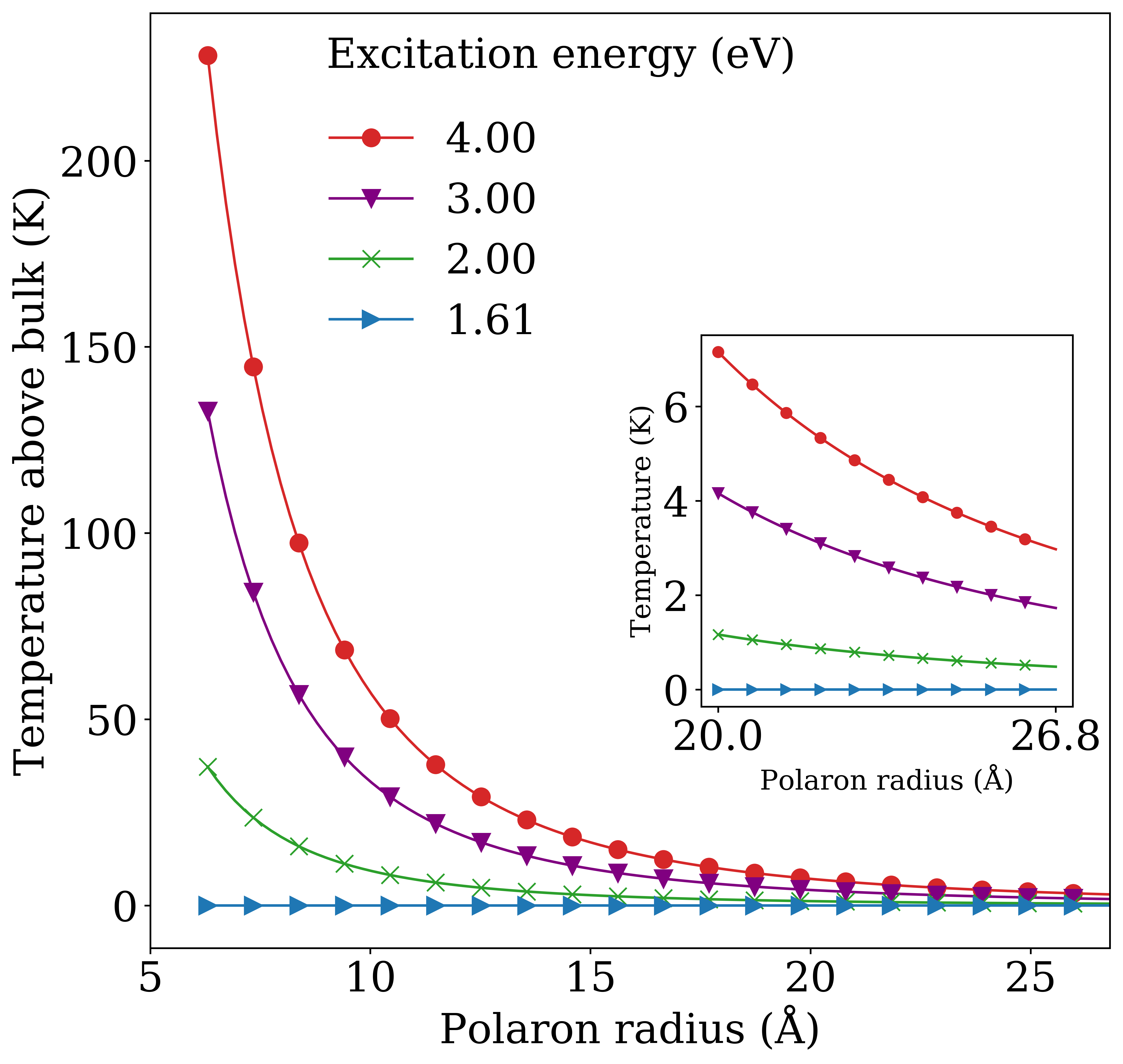}
  \caption{Thermalised polaron temperature in MAPI as a function of polaron
    radius and excitation  energy assuming a bulk value of the heat capacity. 
    The calculated bulk electron polaron radius
    of 26.8 \AA  ~ provides an upper bound for polaron size.
    We take the lattice parameter (6.3 \AA) as a lower bound---below this the
    continuum large polaron approach is not valid. 
    We consider excitation from the bandgap to near-UV. 
    Inset shows detail at larger radii (axes same as main). }
  \label{TemperatureRadius}
\end{figure}

The polaron radius we have calculated is an upper bound: bulk polaron states
are further localised by disorder\cite{Grein1987}. 
The temperature increases with localisation ($T \propto r^{-3}$) as shown in Figure \ref{TemperatureRadius}. 
Point and extended defects (surfaces, interfaces, dislocations, grain
boundaries) may localise polarons further, and so be exposed to local heating
and degradation of the halide perovskite material.  

\textbf{Carrier cooling: initial thermalisation.}
The same force driving polaron formation in MAPI, the dipolar
electron-phonon interaction, will dominate the initial hot carrier
thermalisation, as zone-centre optical phonons are generated. 
The calculated optical-phonon inelastic scattering time is
$\tau=$\SI{0.12}{\pico\second} at \SI{300}{\K}\cite{FrostPolarons2017}. 
The characteristic optical phonon frequency for MAPI is
\SI{2.25}{\tera\hertz}\cite{FrostPolarons2017}, making
a quanta ($E=\hbar\omega$) of this vibration equal to \SI{9.3}{\milli\eV}. 
The thermalisation rate by optical-phonon emission from the polaron is thus
$\frac{\hbar\omega}{\tau}$ = \SI{77.5}{\milli\eV\per\pico\second}. 
This provides an estimate of initial polaron thermalisation.

Energy exchange will proceed until the charge carrier is in thermal equilibrium
with the sub-population of coupled phonons. 
This sub-population will consist of the zone-centre (infrared active) phonon
modes in the near vicinity of the polaron. 
The small size of this population means that the effective specific heat
capacity is reduced, a higher effective polaron temperature will be reached that
predicted from bulk values. 
This occurs on a quantised (per photon) basis, due to the small set of coupled
phonon states in the polaron. 

\textbf{Carrier cooling: heat transfer to the lattice.}
Similar to electrical conductivity, phonon conductivity is limited by scattering events. 
In the bulk, the most frequent is phonon-phonon scattering. 
Due to energy and momentum conservation rules, three-phonon scattering is the lowest-order process. 
We previously\cite{Whalley2016} calculated the three-phonon interaction strengths for
MAPI and found them to be orders of magnitude stronger than for CdTe and GaAs. 
These interactions provide the rates for a stochastic (master equation)
representation of how energy flows microscopically towards equilibrium. 
Direct propagation of this equation with time would provide a microscopic
picture of how the subpopulation of phonon states in a polaron
scatter and cool. 

Here we consider the bulk effect of phonon-phonon scattering. 
The sum of modal contributions, accounting for phonon lifetime, group velocity
and heat capacity, gives the overall thermal conductivity.\cite{Togo2015}
In MAPI, the bulk thermal conductivity from a solution of the Boltzmann
transport equation (in the relaxation time approximation) is extremely low,
\SI{0.05}{\watt\per\metre\per\K} at \SI{300}{\K}.\cite{Whalley2016} 
In contrast, the calculated conductivity for GaAs and CdTe is 
\SI{38}{\watt\per\metre\per\K}
and
\SI{9}{\watt\per\metre\per\K},
respectively. 

To assess the role of the organic cation, a thermal conductivity
calculation was made on CsPbI$_3$ in the cubic perovskite phase. 
Due to the high ($O_h$) symmetry, the computational cost is greatly reduced when compared
against lower symmetry hybrid-halide structures. 
A complication is that the vibrational instability of the cubic \ce{CsPbI3}
structure results in a branch of modes having an imaginary frequency, which is
not considered in the Brillouin zone summations. 
In reality, the room temperature structure of many perovskites is dynamically
cubic\cite{Beecher2016,Bertolotti2017} and such higher-order anharmonicity is
not considered here. 
%
%
%
The calculated thermal conductivity for CsPbI$_3$ is
\SI{0.5}{\watt\per\metre\per\kelvin} at \SI{300}{\K}. 
While still low, it is an order of magnitude greater than
\SI{0.05}{\watt\per\metre\per\kelvin} for MAPI. Kovalsky et
al.\cite{Kovalsky2017} recently measured thermal conductivity in CsPbI$_3$ as
\SI{0.45}{\watt\per\metre\per\kelvin} and in MAPI as
\SI{0.3}{\watt\per\metre\per\kelvin},
with the differences attributed to rotations of \ce{CH3NH3+}.
Additional contributions from electron and ion heat transport, and issues with
sample purity, may explain some disparity between theory and experiment. 

%

\begin{figure}[h]
\centering
  \includegraphics[width=0.95\columnwidth]{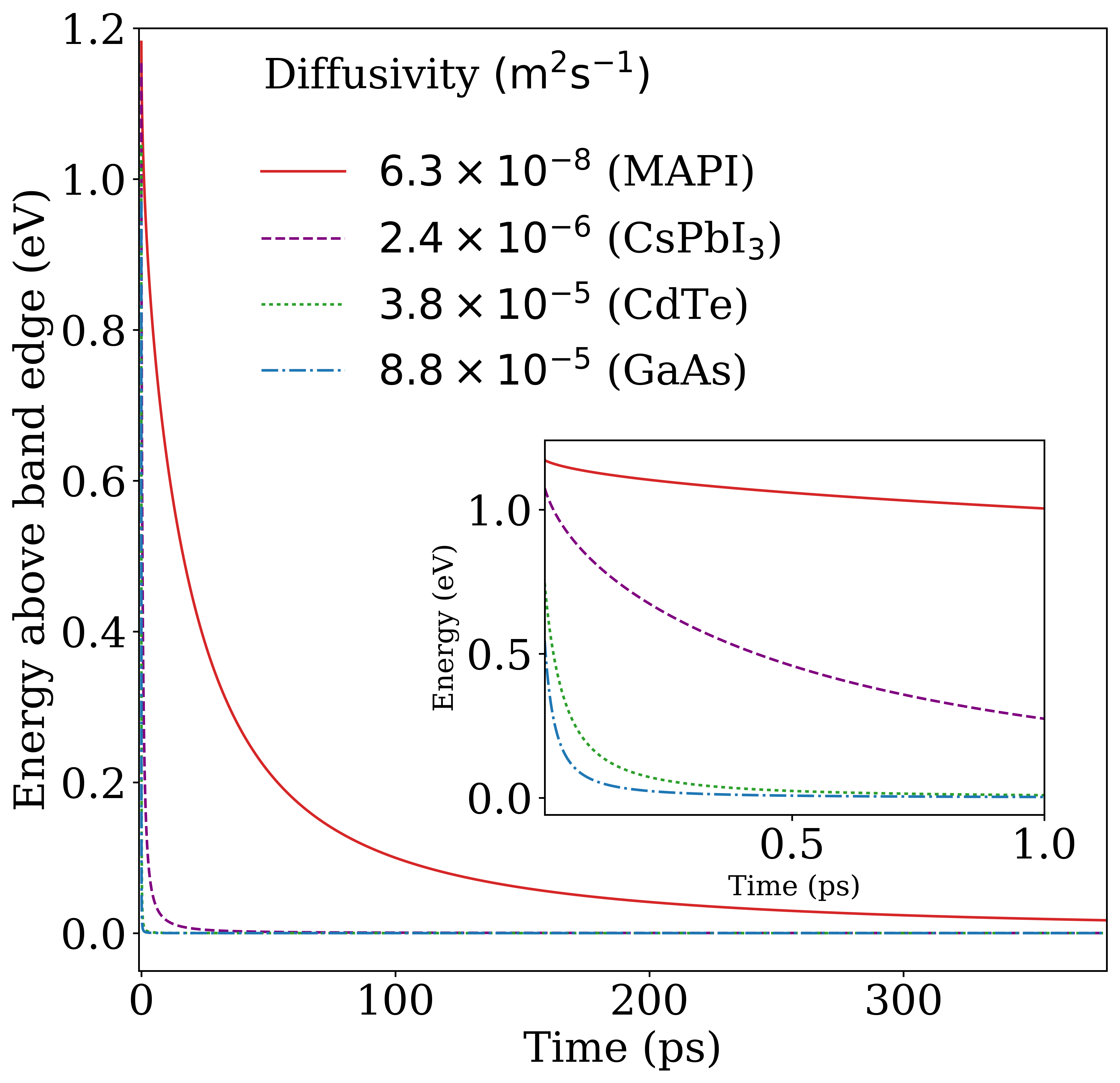}
  \caption{The energy of a large polaron state (starting at 1.2 eV above the conduction band minimum, with a polaron radius of 26.8 \AA) 
  in \ce{CH3NH3PbI3} as a function of time.
  The slow rate is due to the low thermal conductivity in MAPI ($\kappa$ = 0.05Wm$^{-1}$K$^{-1}$). 
  For comparison, we show the behaviour using the thermal conductivities of 
  CdTe ($\kappa$ = 9) and GaAs ($\kappa$ = 38). 
  Heat diffuses from MAPI on the order of 100 ps, whilst for other conductivities 
  the process is much faster, on the order of 100 fs. 
  This is in agreement with reported experimental values: Refs. \onlinecite{Klein2016} (MAPI), \onlinecite{Zhong2017} (CdTe) and \onlinecite{Rosenwaks1993} (GaAs). 
  Note that at short timescales, measurements of hot carrier cooling in MAPI and \ce{CsPbI3}($\kappa$ = 0.5) may appear linear due to the slower exponential decay.}
  \label{TemperatureTime}
\end{figure}

We first consider bulk heat diffusion in the low fluence limit. 
Individual photon quanta are absorbed into isolated hot polarons, cooling by
scattering into phonon modes, which then diffuse away from the polaron. 
Modelling this classically, we can consider the polaron as a hot sphere in
a continuum of ambient temperature material. 
This reduces to a 1-dimensional problem, where the exponent is weighted by the
$r^2$ increasing shell of available states over the surface of the sphere. 
The initial `top hat' heat distribution is convolved with a Gaussian kernel to
give an analytical expression for the evolution of hot carrier energy with time (shown
in Figure \ref{TemperatureTime}).
The rate of cooling is determined by the diffusivity ($D$): 
\begin{equation}
    D= \frac{\kappa}{\rho c_p}
\end{equation} 
where $\kappa$ is the thermal conductivity, $\rho$ is the density and $c_p$ is the specific heat capacity. 
Phonon-phonon cooling in MAPI is on the order of 100 ps. 
This compares well to the observed timescale of slow carrier cooling.
In \ce{CsPbI3} a higher thermal conductivity results in faster polaron cooling on the order of picoseconds.
This effect is stronger than would be expected by na\"ive consideration of the
diffusivity, due to the $D^{\frac{3}{2}}$ scaling of heat conduction from
a point in three dimensions.
%

The phonon-bottleneck is associated with a diminished sub-population of phonon
states, originally envisaged in the gapped density of states present in low
dimensional structures\cite{Bockelmann1990}. 
A reduced population of vibrational states strongly couple to the
charge carrier in the polaron state. 
These are the infrared-active phonon modes, identified in lattice
dynamic studies\cite{Brivio2015} as octahedral distortion modes
of the \ce{PbI3-} framework.
Such distortions for illuminated MAPI have been 
observed using a time-dependent local structure analysis,\cite{wu2017}
and further signatures of polaron formation observed in the bromide compound.\cite{miyata2017}

At low fluence, the sub-population of polaron phonon modes will thermalise the carrier to
a higher temperature than expected for the lattice.
The strong phonon-phonon scattering introduces a \SI{100}{\ps} time constant
for the bulk flow of thermal energy out of an isolated polaron, which broadly
agrees with the observed time constants. 
A more detailed understanding of this out-of-equilibrium energy flow could be
made by studying the microscopic phonon-phonon scattering cross-sections, and
considering the modal heat capacities.

Additionally, at high fluence, the polaron states overlap, 
so diffusion of phonons away from the polaron simply results in reheating other
polarons.  
There is no thermal gradient to drive diffusion. 
In both cases, eventual cooling will proceed by scattering into other (non
electron-phonon coupled) phonon modes.

In summary, we have shown how effective mass theories of excitons and
polarons---informed by first-principles calculations---can be combined to
describe the physical processes behind the slow hot-carrier cooling rates
observed for halide perovskites.
From an interpretation of the density at which the polarons start to
overlap, we indicate that significant changes in the photophysics should occur
when $n \ge$ \SI{e18}{\per\cm\cubed}. 
This corresponds to the observed transition region between low-fluence `high
energy photoluminescence' and high-fluence `hot-phonon bottleneck'
regimes\cite{Bretschneider2017}.
We have underlined the unusual electronic structure of hybrid halide
perovskites, possessing a second conduction band at +\SI{2.5}{\eV}
above the valence band, and therefore caution careful
interpretation of photophysics data when pumping with photon energies $>$\SI{2.5}{\eV}.
Finally, we calculated a higher thermal conductivity in the inorganic
pervoskite, compared to the organic-cation hybrid perovskite. 
This can help explain the lack of hot-carrier photoluminescence in the
Cs-based material\cite{Zhu2016}, and emphasises the phonon scattering `rattler'
role of the organic cation in limiting thermal dissipation of hot carrier energy.

\textbf{ACKNOWLEDGMENTS}

Via our membership of the UK's HEC Materials Chemistry Consortium, which is funded by EPSRC (EP/L000202), this work used the ARCHER UK National Supercomputing Service (http://www.archer.ac.uk).
This work was funded by the EPSRC (grant numbers EP/L01551X/1, EP/L000202, and EP/K016288/1), the Royal Society, and the ERC (grant no. 277757). 


\textbf{ASSOCIATED CONTENT}

Data files and Jupyter notebooks outlining the calculation steps are available as a repository on GitHub at https://github.com/WMD-group/hot-carrier-cooling.

\bibliography{MAPI-Thermal-Pathways}

\end{document}